# De-aberration for noninvasive transcranial photoacoustic computed tomography through an adult human skull

Yousuf Aborahama[1,†], Karteekeya Sastry[1,2,†], Manxiu Cui[1,†], Yang Zhang[1], Yilin Luo[1], Rui Cao[1], Geng Ku[1], Jigmi Basumatary[1], Junhao Zhu[1], Siying Kong[1], Lihong V. Wang[1,2,*]

[1]Caltech Optical Imaging Laboratory, Andrew and Peggy Cherng Department of Medical Engineering, California Institute of Technology, 1200 East California Boulevard, Pasadena, CA 91125, USA.

[2]Caltech Optical Imaging Laboratory, Department of Electrical Engineering, California Institute of Technology, 1200 East California Boulevard, Pasadena, CA 91125, USA.

[†]These authors contributed equally to this work: Yousuf Aborahama, Karteekeya Sastry, Manxiu Cui.

[*]Corresponding author: Lihong V. Wang (LVW@caltech.edu).

## Abstract

Noninvasive transcranial photoacoustic computed tomography (PACT) of the human brain, despite its clinical potential as a complementary technology to functional MRI, remains impeded by the acoustic distortion induced by the human skull. The distortion, which is attributed to the markedly different material properties of the skull relative to soft tissue, results in heavily aberrated PACT images—a problem that has remained unsolved for the past two decades. Herein, we report the first successful experimental demonstration of the de-aberration of PACT images through an ex-vivo adult human skull using a homogeneous elastic model for the skull. Using only the geometry, position, and orientation of the skull, we faithfully de-aberrate the PACT images of light-absorbing phantoms acquired through an ex-vivo human skull for different levels of phantom complexity and positions. We also demonstrate the generality of our results by attaining a similar extent of de-aberration through a second ex-vivo human skull. Our work addresses the longstanding challenge of skull-induced aberrations in transcranial PACT and advances the field towards unlocking the full potential of transcranial human brain PACT.

## Introduction

Photoacoustic computed tomography (PACT) is an emerging technique that enables molecule-specific optical absorption imaging at sub-millimeter resolution and centimeter-scale depth (*1*). It is based on the generation of acoustic waves upon the absorption of light by tissue, namely, the photoacoustic (PA) effect (*2*). Owing to the approximately homogeneous acoustic properties of soft tissues, the generated initial pressure map in tissue, which is proportional to the optical absorption coefficient distribution, can be readily reconstructed from the recorded PA signals around the target using the universal backprojection (UBP) method (*3*). Hence, PACT has been used for imaging several parts of the human body, such as the breast (*4–6*), extremities (*7, 8*), and neck (*9, 10*).

PACT of the human brain has also been investigated (*11*, *12*) since it holds several advantages as a neuroimaging modality compared to the widely-used functional magnetic resonance imaging (fMRI). PACT is directly sensitive to both oxy- and deoxy-hemoglobin linearly with a low tissue background, thus offering multiple functional contrasts as opposed to fMRI, which is only indirectly sensitive to changes in deoxy-hemoglobin in a nonlinear manner. It is more portable, more space-efficient, more open, less noisy, less expensive, and cheaper to maintain than fMRI. Further, due to its magnet-free operation, it can be used for subjects with implants that are incompatible with fMRI. Recently, PACT of human brain vasculature and function was demonstrated in the special population of hemicraniectomy patients in whom a section of the skull is surgically removed to release elevated brain pressure (*13*). It was validated with blood oxygen level-dependent (BOLD) fMRI at the highest clinically allowed magnetic field strength, thus demonstrating its clinical potential. However, for this technology to be applicable to the general adult human population, the challenge posed by the human skull needs to be overcome.

The mechanical properties of the human skull differ significantly from that of soft tissue (*11*). Unlike soft tissue, the skull supports both compression and shear waves, thus resulting in mode conversion at the soft tissue-skull interface. The average compression wave speed and the density of the skull are also approximately twice that of soft tissue. Moreover, the skull is a dispersive medium and induces frequency-dependent attenuation. Thus, the waves passing through the skull are severely distorted, which causes the resulting PA images, reconstructed using the UBP method, to be highly aberrated. This distortion is also detrimental to human brain functional imaging since it degrades the strength of the detected functional signals and the accuracy of localization of functional regions, as shown in this preliminary study (*14*). To overcome this problem, the properties of the skull need to be incorporated into the image reconstruction method.

Despite numerous attempts over the last two decades (*15–23*), a conclusive experimental demonstration of de-aberration in transcranial PA images remains elusive. A layered back-projection method (*19*) was developed for transcranial PACT by extending UBP to a piecewise homogeneous medium. However, this method cannot account for reverberations within the skull and reflections of waves arising from absorbers outside the skull. Further, it did not accurately model the brain-skull and skull-scalp interfaces, which resulted in only a partial correction of the skull-induced aberrations. The memory effect of skull-induced distortions in PA signals arising within a small "isoplanatic" intracranial region was exploited to de-aberrate images of a collection of point-absorbing targets through a piece of excised skull obtained from the fronto-temporal region (*21*). However, it relied on the invasive prior measurement of a point source signal through the skull. Several other approaches have been proposed in the literature, but they have only been demonstrated with a simian skull (*15*, *16*) or an acrylic globe (*17*, *18*), which induce significantly less distortion than an adult human skull.

Here, for the first time, we demonstrate the de-aberration of PACT images of light-absorbing phantoms of vasculature through an ex-vivo adult human skull using a homogeneous elastic model for the skull. Using only the geometry, position, and orientation of the skull, we can achieve high-fidelity aberration correction in the phantom images, in terms of the recovered features. Our work addresses the longstanding problem of skull aberration in transcranial PACT and takes us one step closer to achieving the goal of human brain PACT.

**Results**

We imaged light-absorbing phantoms through an ex-vivo adult human skull using a three-dimensional (3D) PACT system similar to the one described here (*24*). We placed the phantoms close to the inner surface of the ex-vivo skull to mimic cortical blood vessels. We took special care to remove any trapped air within the skull trabeculae (see Methods). We use two illumination approaches—internal and external illumination, respectively. The internal illumination leads to an improved signal-to-noise ratio (SNR) in the PA signals, thus allowing us to effectively test our de-aberration method, whereas the external illumination is used to demonstrate our method in noninvasive transcranial PACT. The external illumination is also used to acquire a PA image of the fiducial markers on the skull, which is used to co-register the X-ray computed tomography (CT) image of the skull with the PACT frame of reference.

A schematic of the experimental setup is shown in Fig. 1A, which illustrates the skull, the imaging target, the two illumination approaches, and the hemispherical ultrasonic detection surface, which was obtained by rotating four arc-shaped ultrasonic transducer arrays. PA waves originate from the imaging target and are aberrated by the skull before reaching the ultrasonic detectors. Reconstructing an image without accounting for the skull leads to an aberrated PA image, as shown in the top row in Fig. 1B. On the other hand, an image reconstruction method that incorporates the skull model can effectively refocus the aberrated waves, thus resulting in a de-aberrated PA image.

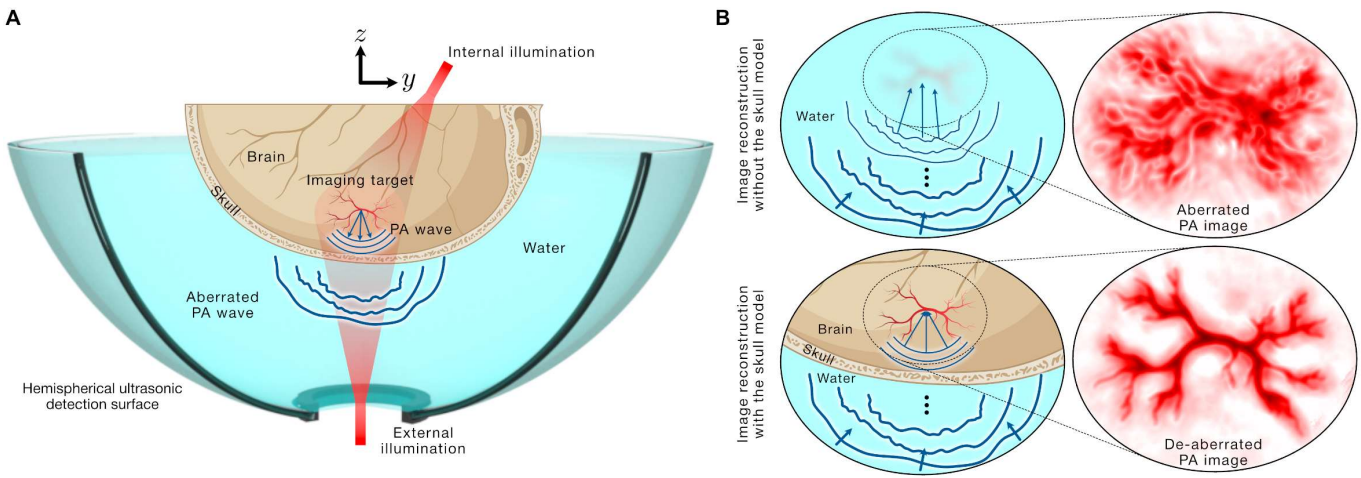

**Fig. 1| Schematic of the experimental setup and the image reconstruction scheme. A**, 3D PACT system comprising a hemispherical ultrasonic detection surface (obtained by rotating four arc-shaped ultrasonic transducer arrays) filled with water. The ex-vivo skull is placed within the imaging field of view. Internal illumination is used to improve the SNR in this experiment, which allows us to effectively test our de-aberration method. External illumination is used to acquire an image of the skull fiducial markers for co-registration and to demonstrate our method in noninvasive transcranial PACT. The PA wave generated from the imaging target gets aberrated by the skull before reaching the ultrasonic detection surface. **B**, Illustration of image reconstruction with and without skull modeling. (Top) Reconstructing an image of the target transcranially without accounting for the skull leads to an aberrated PA image. (Bottom) Incorporating the effect of the skull in the reconstruction method leads to the correction of the aberrated wavefronts, thus resulting in a de-aberrated PA image.

We reconstruct the images of several phantoms acquired in the absence and presence of the skull using UBP, and acquired in the presence of the skull using our approach and present them along with their respective ground truth (i.e., their photographs) in Fig. 2A. The images are presented as maximum amplitude projections (MAPs) of the relevant slices in the respective reconstructed volume. Phantom 1 is made of polylactic acid (PLA), whereas phantoms 2 and 3 are made of blood-filled tubes embedded in agarose. We see in the case of all three phantoms that the images acquired in the presence of the skull, reconstructed using UBP, are severely aberrated and hardly capture any of the features of the respective targets. Remarkably, for the same phantoms, the de-aberrated transcranial images, shown in the right-most column in Fig. 2A, recover nearly all the features that can be seen in the respective ground truth images as well as the UBP images of the phantoms acquired in the absence of the skull. Further, we extract two line profiles from the images of each phantom and plot them in Fig. 2B, which shows that the de-aberrated phantom images closely resemble the images acquired in the absence of the skull. A detailed description of the preprocessing and image reconstruction scheme is presented in Methods.

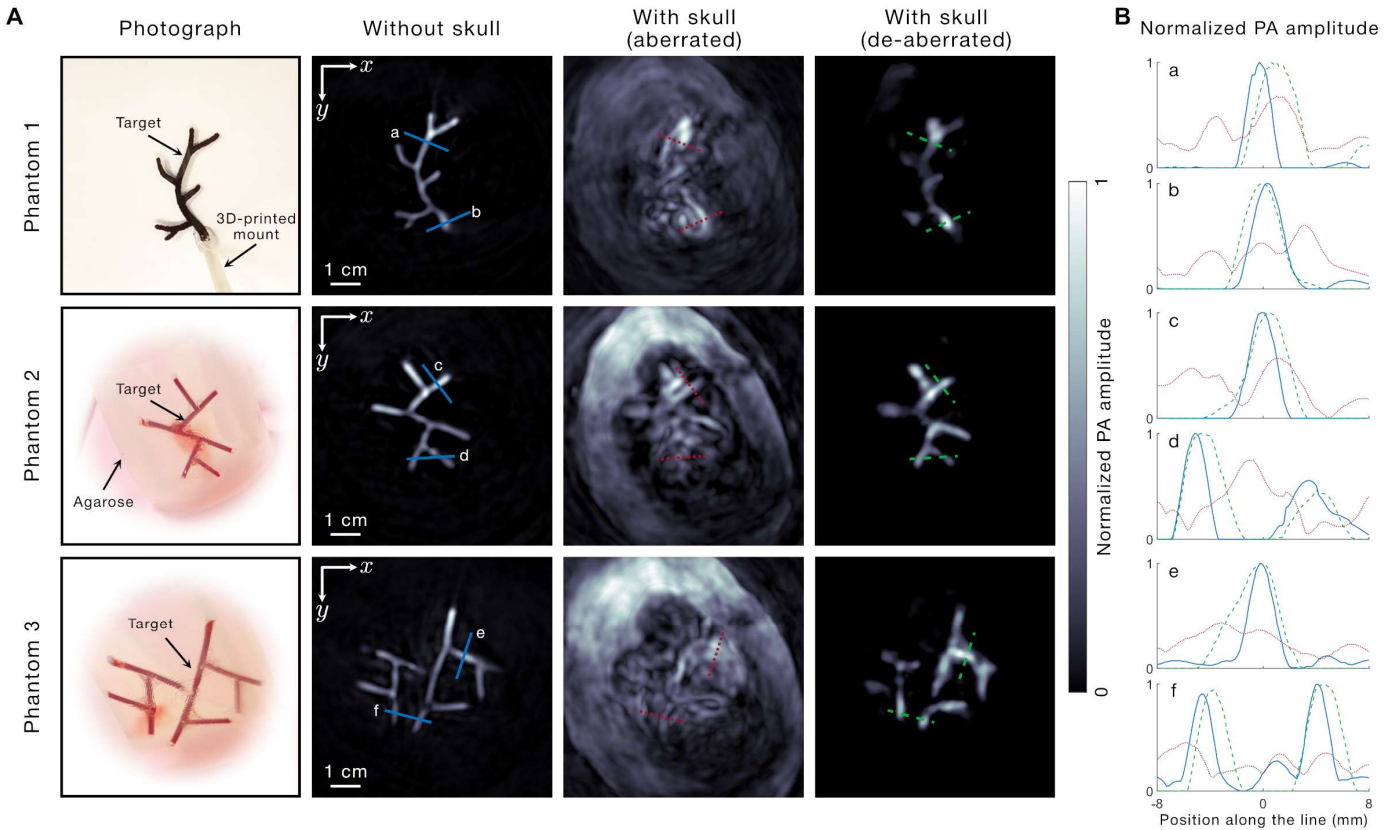

**Fig. 2| Experimental demonstration of skull de-aberration in PACT. A**, Images of three light-absorbing phantoms are acquired with and without the presence of the ex-vivo skull. The photographs of the phantoms are shown in the left-most column as the ground truths. The UBP images of the phantoms acquired in the absence and presence of the skull are shown in the second and third columns, respectively. Finally, the de-aberrated images acquired through the skull are shown in the right-most column. The de-aberrated images remarkably recover most of the features that are seen in the respective ground truths (i.e., the photographs and the UBP images acquired without the skull), whereas the UBP images acquired in the presence of the skull hardly reveal any of the features. **B**, Plots of the line profiles extracted from the images of the phantoms (two profiles for each phantom). The colors and line styles of the respective profiles match the plotted lines in **A**.

The images in Fig. 2 were obtained by internally illuminating the targets to ensure high SNR. However, we can also de-aberrate noninvasively through the skull using external illumination as shown in Fig. 3. In Fig. 3A, we present the UBP image of a phantom (made of black wires) in the absence of the skull along with its photograph. Then, we show the aberrated and de-aberrated images obtained using internal and external illuminations, respectively, thus demonstrating the noninvasive de-aberration of the phantom image through the skull. We also extract line profiles from the phantom images in Fig. 3B, which show that the de-aberrated images obtained using external and internal illuminations, respectively, are comparable to the image obtained in the absence of the skull, in terms of the resolution and recovered phantom position. An interesting observation from Fig. 3A is that while the phantom is visible, albeit distorted, in the aberrated image obtained using internal illumination, it is hardly discernible from the background in the aberrated image obtained using external illumination. This is because of the background generated by the skull, which is much stronger in the external illumination case than in the internal illumination one. This deterministic interference from the skull is hugely detrimental to ex-vivo transcranial PACT—much more than random noise which can be mitigated by averaging—since it dominates the weaker signals arising from within the skull. In the in-vivo case, a similar deterministic background is generated by the strong signal generated in the scalp and its reflection from the skull. However, as we have shown, we can partially overcome the background by correcting the skull-induced aberrations, thereby focusing the target inside the skull.

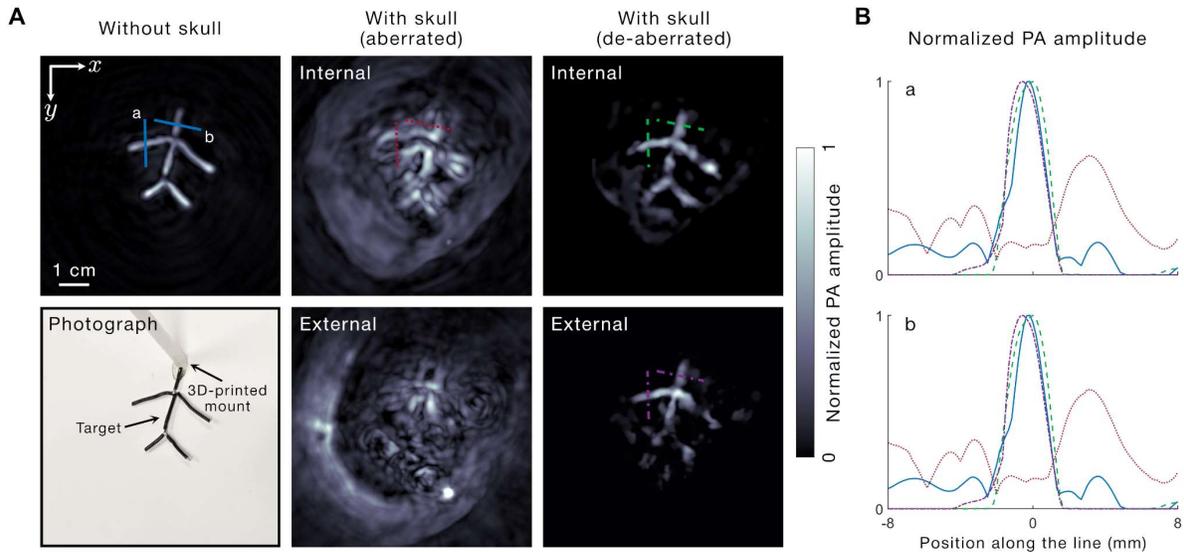

**Fig. 3| De-aberration in noninvasive imaging through an ex-vivo adult human skull. A**, UBP image in the absence of the skull, photograph, UBP (aberrated) images in the presence of the skull obtained with internal and external illumination, and the de-aberrated images in the presence of the skull with internal and external illumination, respectively. **B**, Line profiles extracted from the images in **A**.

We also acquire the images of a phantom (made of black wires) through the skull at different positions along the $z$ direction using internal illumination. The photograph and the UBP image in the absence of the skull, respectively, of the phantom are shown in Fig. 4A. In Fig. 4B, we show the cross-sections (in the $x$-$z$ plane) of the de-aberrated PACT images of the phantoms acquired at the three positions (within 16 mm along the $z$ direction) overlaid on a cross-section of the registered X-ray CT image of the ex-vivo skull. This shows the positions of the phantoms relative to the ex-vivo skull. Further, as evidenced in Fig. 4B, the average thickness of the relevant portion of the ex-vivo skull is roughly 6.5 mm. At each position, we reconstruct the aberrated (UBP) and de-aberrated images and present them in Figs. 4C and 4D, respectively. These images demonstrate that the de-aberration is effective for different depths relative to the skull.

To further demonstrate the fidelity of our de-aberration approach, we acquire images of phantoms through another ex-vivo skull (using external illumination). While the shape of the first ex-vivo skull was obtained using X-ray CT, we extracted the shape of the second ex-vivo skull using magnetic resonance (MR) imaging (see Methods for details), which does not involve any ionizing radiation like X-ray CT. Since our objective is to demonstrate the de-aberration of the phantom images, we isolate the transcranial phantom signal by subtracting a skull-only signal from the total acquired signal, which reduces the interference from the skull background, akin to the internal illumination images. We present the aberrated (UBP) and de-aberrated images of two phantoms obtained without the skull background in the first two columns of Fig. 4E. These images show that we can de-aberrate phantom images reliably through either of the ex-vivo skulls and demonstrate the robustness of this approach. We also show the de-aberrated images of the two phantoms obtained with the skull background in the third column, which, like Fig. 3B, exhibit more background than those without the skull background, but still reveal the respective phantom features accurately. A cross-section (in the $x$-$z$ plane) of the de-aberrated image (without the skull background) of phantom 1 in Fig. 4E overlaid on the cross-section of the MR image of the second ex-vivo skull is shown in Fig. 4F. The stark improvement in the de-aberrated images across different levels of phantom complexity, different phantom positions, and even different skulls shows the reliability and generality of our reconstruction scheme.

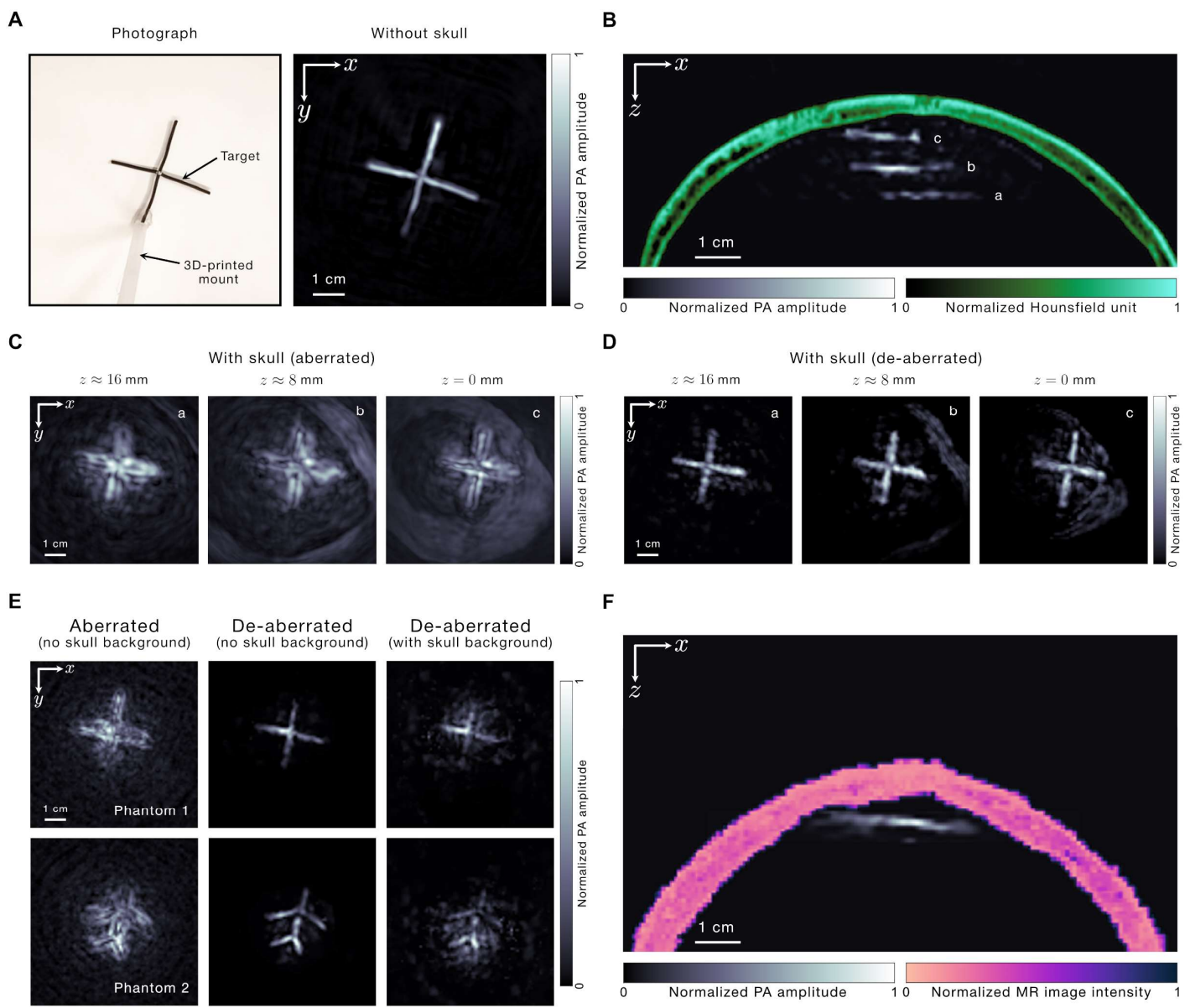

**Fig. 4| De-aberration at different phantom positions relative to the skull and for a second ex-vivo skull.**
**A**, Photograph and UBP image in the absence of the skull, respectively, of a phantom. **B**, Cross-sections of the de-aberrated PACT images of the phantom acquired at different positions relative to the skull overlaid on a cross-section of the registered X-ray CT image of the ex-vivo skull. Phantom position c is closest to the skull and is set as the $z = 0$ plane. **C-D**, Aberrated (**C**) and de-aberrated (**D**) images of the phantom, respectively, acquired in the presence of the skull, at different $z$ positions relative to the skull. The aberrated images in **C** are reconstructed using UBP. **E**, Left and middle columns: Aberrated and de-aberrated images of two phantoms without the skull background, respectively, acquired in the presence of the second ex-vivo skull. Right column: De-aberrated images of the two phantoms with the skull background, acquired in the presence of the second ex-vivo skull. **F**, Cross-section of the de-aberrated image of phantom 1 in **E** (without the skull background) overlaid on the cross-section of the MR image of the second ex-vivo skull.

The performance of any model-based image reconstruction approach depends on the accuracy of the underlying model. The accuracy of our model depends on the correctness of the position, orientation, and properties of the skull. To test the sensitivity of our method to errors in these properties, we perturb each of these quantities and study their effect on the de-aberration in Fig. 5. In Fig. 5A, we show the photograph of a phantom (made of PLA) along with its UBP images in the absence and presence of the skull, respectively, and its de-aberrated (using the iterative reconstruction method described in Methods) and adjoint-reconstructed (see Methods) images in the presence of the skull, respectively. For evaluating the effect of model-mismatch, we consider the adjoint-reconstructed image under different scenarios. First, we evaluate the effect of ignoring shear waves by modeling the skull as a medium that only supports compression waves (i.e., an acoustic model) and present this image in Fig. 5B as a model perturbation. The noticeable deterioration in the reconstructed image quality shows the importance of modeling the shear component. To study the impact of errors in the skull position in the elastic model on de-aberration, we translate the skull by a maximum of around 1 cm in different directions and pick the image that has the least correlation coefficient (CC) with the image obtained without translation. We present this image (CC: 0.60) in Fig. 5B. Since a shift in the image position negatively affects the CC, we consider a "sliding correlation", which is computed as the maximum of the normalized cross-correlation between the two images. We perturb the orientation of the skull by rotating it about the coordinate axes by ±10º. Once again, we pick the image with the least correlation with the image obtained without rotation and present it (CC: 0.62) in Fig. 5B. We also change the assigned compression and shear speed in the skull by ±10% and present the image with the least correlation (CC: 0.71) in Fig. 5B. Notably, although the skull position, orientation, and speed of sound perturbation result in a degradation in the image quality, many of the features of the phantom are still visible. This shows that while the accuracy of the skull parameters is necessary for optimal de-aberration, small errors in these parameters may still result in comparable images. Finally, we plot the correlation coefficients (with respect to the adjoint-reconstructed image in Fig. 5A) for different assigned values of compression and shear speeds within the skull in Fig. 5C, thus showing that changes in the shear speed seem to affect the image quality more than changes in the compression speed. Since the correlation coefficient plots for the position and orientation perturbation were not elucidative and they depend on the phantom placement relative to the skull, making them less generalizable, we have not included them here.

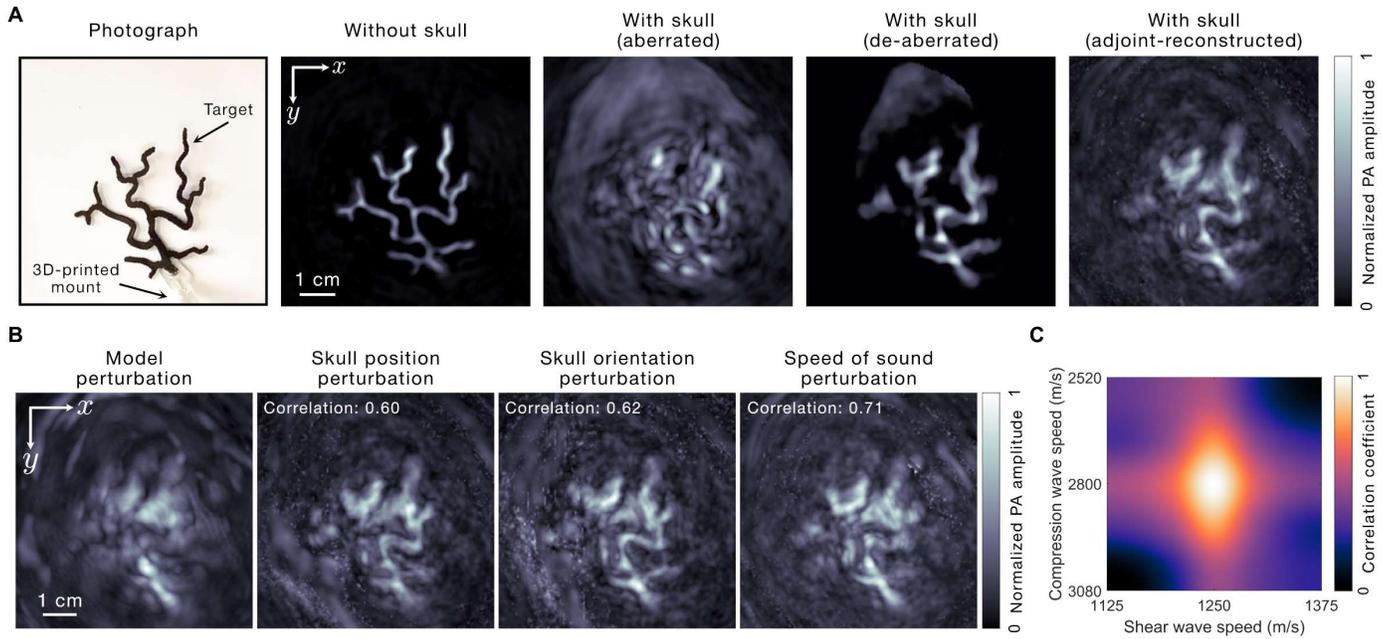

**Fig. 5| Impact of model-mismatch on de-aberration. A**, Photograph of a phantom along with its UBP images in the absence and presence of the skull, respectively, and its de-aberrated and adjoint-reconstructed images in the presence of the skull, respectively. **B**, Adjoint-reconstructed images under different conditions of model-mismatch. The leftmost image (model perturbation) is reconstructed by modeling the skull as a medium that only supports compression waves (i.e., an acoustic model) and it shows the importance of modeling the shear component. The second, third, and fourth images are reconstructed using the elastic model by perturbing the skull position (CC: 0.60), orientation (CC: 0.62), and compression and shear speeds (CC: 0.71). These three images do not show a severe degradation of image quality and demonstrate that our method is robust to errors in these parameters. **C**, Plot of the correlation coefficients (with respect to the adjoint-reconstructed image in **A**) of the images reconstructed with different compression and shear speeds within the skull. It shows that changes in the shear speed seemingly affect the image quality more than the changes in the compression speed.

## Discussion

Here, we presented, for the first time, de-aberrated PACT images of phantoms through an ex-vivo skull, which addresses a long-standing challenge faced in PACT. Using a homogeneous elastic model for the skull, we achieved successful de-aberration, in terms of feature recovery, for phantoms with different levels of complexity. We demonstrated the same degree of de-aberration noninvasively, i.e., by illuminating the phantoms from outside the skull. The presented approach can be utilized for transcranial imaging at different depths as shown from the results for different phantom positions relative to the skull. We have also demonstrated the generality of our method by showing a similar extent of aberration correction through a second ex-vivo adult human skull. Finally, to ensure the usability of the proposed method in a practical setting, we investigated the sensitivity of our de-aberration approach to various model parameters. While modeling the shear component was found to be crucial to de-aberration, the model was resilient to small errors in the skull position, orientation, and speeds, with the shear speed affecting the performance more than the compression speed. These results convincingly demonstrate the correction of skull-induced aberrations in transcranial PACT, thus establishing its potential as a complementary high-resolution neuroimaging modality to fMRI.

Our approach holds several technical and practical advantages over existing methods in the literature such as the layered back-projection (*19*) method (LUBP) and the memory effect-based method (*21*). Unlike the LUBP method, which ignores the polarization of shear waves, reflections from and within the skull, and uses multiple approximations to handle the wave propagation, our approach follows a full-wave modeling scheme, which handles the fluid-solid interface more accurately, is adaptable to inhomogeneous models of the skull, and can be used in an iterative framework that can compensate for modeling errors by leveraging prior information. The memory effect-based approach, on the other hand, requires an invasive calibration measurement and is valid only within a small region, thus necessitating several invasive calibration measurements for imaging a large region. In contrast, our approach, which only requires the geometry, position, and orientation of the skull, along with representative values for the skull properties, is much more clinically viable. Finally, although the isotropic elastic wave equation has been proposed before to model the skull (*17*, *18*), our work demonstrates its first application for successful aberration correction through an adult human skull in transcranial PACT.

The geometries of the two ex-vivo skulls used in this work were obtained using X-ray CT and MR imaging, respectively, and fiducial markers on each skull were used to register the skull in the PACT frame of reference. Even in the in-vivo case, X-ray CT can be used to obtain the shape of the skull. To register the skull in the PACT frame of reference, an additional MR angiography scan of the head can be performed. The PACT frame of reference can be registered with the MR frame of reference using a previously established framework (*13*), which uses superficial scalp arteries that appear in both the PACT and MR angiography images of the head. The MR frame of reference can then be registered with the X-ray CT image using multimodal medical image registration algorithms (*25*). Alternatively, certain MR imaging sequences may be used to directly obtain the skull geometry (*26–28*).

PACT of the human brain, coupled with our de-aberration approach, has attractive clinical potential. Previous studies (*11*, *13*) have concluded that sufficient SNR is achievable in in-vivo transcranial PACT with sufficient illumination (within the safety limit) and well-designed ultrasound detector arrays. In the current work, PACT may be a complementary technique to X-ray CT or MR imaging since one of these modalities is first used to estimate the skull parameters. Nonetheless, PACT can be subsequently used to monitor brain function or other clinically relevant information and to aid in the diagnosis and management of brain diseases such as traumatic injuries, tumors, and strokes (*29–31*). This is particularly beneficial since PACT does not involve exposure to harmful ionizing radiation or administration of contrast agents. PACT is well suited for patients with either fMRI-incompatible implants or claustrophobia or patients who need frequent bedside monitoring at a lower cost. In the future, a combined PA-ultrasound imaging system is envisioned, in which ultrasound imaging is used to image the properties and shape of the skull.

A few aspects can be further improved. Firstly, we can use a heterogeneous model for the skull. Although our approach is readily extendable to the heterogeneous case, it is challenging to estimate the elastic properties of the skull accurately. While some literature exists on mapping the Hounsfield units in X-ray CT images to the compression and shear wave speeds (*32–35*), a consensus on the optimal solution still does not exist (*36*), especially for the shear wave speed. Estimating the skull properties from an MR image is another important direction since it obviates the need for an X-ray CT image, which involves ionizing radiation. Second, we did not consider the effects of dispersion and frequency-dependent attenuation within the skull in our model (*11*). Incorporating these effects will improve the accuracy of the skull model, thus resulting in better de-aberration and a higher resolution. Finally, we need to study the effect of the PA signals arising outside the skull (e.g., hair follicles and melanin in the human scalp), which get reflected from the skull and potentially corrupt the cortical signals. Accounting for these reflections is the next major challenge in transcranial PACT and will be the focus of our future work.

## Methods

### Sample preparation

The two ex-vivo human skulls (Skull Unlimited International Inc.) used in this study were donated by an 83-year-old male and a 51-year old female, respectively. We immersed each ex-vivo skull in water for more than 2 hours before the experiment and used a vacuum pump to remove any trapped air in the skull trabeculae. The MR image of the second ex-vivo skull, submerged in deionized water to create a negative contrast, was acquired using a 3T Siemens Prisma.Fit scanner with a 32-channel head receive coil at the Caltech Brain Imaging Center. A multi-echo rapid gradient echo (MERAGE) sequence was used with a root mean square (RMS) echo combination (repetition time = 2210 ms, RAGE echo time = 1.6, 3.5, 5.3, and 7.1 ms, flip angle = 8º, in-plane generalized autocalibrating partially parallel acquisitions (GRAPPA) acceleration factor = 2, pixel bandwidth = 700 Hz, 0.9 × 0.9 × 0.9 $mm^3$ isotropic voxel, 3D distortion correction, total image acquisition time = 247 seconds). The phantoms used in this paper were either made of black wires, were 3D-printed using polylactic acid, or were made of blood-filled tubes embedded in agarose. The blood tube phantoms consisted of a micro-renathane tubular structure (0.066" inner diameter, Braintree Scientific) filled with bovine blood (B-C8080, defibrinated, QuadFive) that was sealed from all ends using hot melt adhesive and embedded in 2.5% agarose (A9414, low gelling temperature, MilliporeSigma). Custom-designed 3D-printed mounts were used to ensure the repeatability of the phantom position.

**Data processing and image reconstruction**

The acquired PA signals were low-pass filtered (*37*) using a second-order Butterworth filter with a cutoff frequency of 0.5 MHz. The Kabsch algorithm (*38*) was used to estimate a rigid transformation between the X-ray CT (or MR) image of the skull and the PA frame of reference. The transformed CT (or MR) image was then binarized and the holes within the binarized skull were filled in using morphological operations (*39*). It was then treated as a homogeneous elastic medium. The background medium was water. Representative values, found in the literature (*11*), were used for the density, compression wave speed, and shear wave speed within the skull. The sound speed in water was estimated from its temperature measurement (*40*).

We used the isotropic elastic wave equation to model wave propagation in the skull, as given by the following initial value problem (*41*) (written in index notation (*42*)):

$$\rho \partial_t v_i = \partial_j \sigma_{ij},$$
$$\partial_t \sigma_{ij} = \lambda \delta_{ij} \partial_k v_k + \mu(\partial_i v_j + \partial_j v_i), \text{ and} \tag{1}$$
$$\sigma_{ij}|_{t=0} = -p_0 \delta_{ij}, \quad v_i|_{t=0} = 0,$$

where $\sigma_{ij}$ is the induced stress tensor at time $t \geq 0$, $v_i$ is the velocity vector, $\partial_i \equiv \partial/\partial x_i$ ($x_i$ denotes the $i^{\text{th}}$ Cartesian position coordinate for $i \in \{1,2,3\}$), $\partial_t$ denotes a temporal derivative, $\delta_{ij}$ is the Kronecker delta, $p_0$ is the initial pressure, $\rho$ is the mass density, and $\lambda$ and $\mu$ are the Lamé parameters.

The images are reconstructed by solving the following regularized least squares problem,

$$\hat{p}_0 = \underset{p_0 \in \mathbb{R}^N_{\geq 0}}{\operatorname{argmin}} \frac{1}{2} \|SAp_0 - d\|^2 + R(p_0), \tag{2}$$

where $S: \mathbb{R}^{N_x \times N_t} \to \mathbb{R}^{N_m \times N_t}$ is a sampling operator that maps the pressure at $N_x$ positions and $N_t$ time instances to the $N_m$ ultrasonic transducers' measurements, $A: \mathbb{R}^N \to \mathbb{R}^{N_x \times N_t}$ is a discrete approximation for the operator that solves the initial value problem in Eq. (1), $N$ is the number of 3D positions at which the initial pressure is being solved for, $\hat{p}_0$ is the estimated initial pressure distribution, $\mathbb{R}^N_{\geq 0}$ is the set of $N$-dimensional real vectors with non-negative entries, $d \in \mathbb{R}^{N_m \times N_t}$ is the data measured by the transducers, and $R: \mathbb{R}^N \to \mathbb{R}$ is the regularizing function, chosen here to be a linear combination of the $L_1$-norm and the isotropic total variation (TV) semi-norm (*43*). We solve the above optimization problem using an accelerated proximal gradient method (*43–45*), where the step size can be estimated using power iterations (*46*). The adjoint-reconstructed images in Fig. 4 are reconstructed by computing the action of the adjoint of $SA$ on $d$, which is equivalent to performing a single iteration in the absence of any regularizers. At the gradient step, the actions of the operator $SA$ and its adjoint, respectively, are evaluated using the pseudo-spectral time-domain-based k-Wave toolbox (*47*). The spatial discretization was 0.5 mm, which is roughly one-sixth of the wavelength of sound in water at the maximum frequency of consideration. The temporal step size was 50 ns and the computational grid was terminated with a perfectly matched layer to avoid spurious reflections from the edges of the domain. The density, compressional wave speed, and shear wave speed of the first ex-vivo skull were set to 1850 kg/m$^3$, 2800 m/s, and 1250 m/s, respectively, and of the second ex-vivo skull were set to to 1850 kg/m$^3$, 2800 m/s, and 1400 m/s, respectively. The weights of the two regularizers were tuned for the different illumination conditions (internal and external illuminations), different imaging target types (black wires, PLA, and blood tubes), and different skulls. This is necessary since the regularizers are sensitive to various factors such as the signal-to-interference ratio and overall signal scale which varies under these different conditions. The optimization was terminated after 10 iterations. The data processing and image reconstruction were carried out in MATLAB R2023b running on a Windows 11 Pro 23H2 computer with an Intel Core i9-13900KS at 3.2 GHz, 4×48 GB DDR5 RAM at 5200 MHz, and an NVIDIA GeForce RTX 4090 GPU.

**Acknowledgments:**

We thank Dr. Michael Tyszka from the Caltech Brain Imaging Center for his help in acquiring the MR image of the skull.

**Funding:** This work was sponsored by the United States National Institutes of Health (NIH) grants U01 EB029823 (BRAIN Initiative), R35 CA220436 (Outstanding Investigator Award), and R01 CA282505.

**Author contributions:** L.V.W. conceived the project. Y.A, K.S., M.C., and L.V.W. designed the study. Y.A. and K.S. developed and optimized the data processing and image reconstruction platform, analyzed the data, and generated the results. Y.Z., M.C., G.K., J.Z., and S.K. built the experimental setup. M.C., Y.Z., K.S., J.B., J.Z., and S.K. performed the experiment and collected the data. Y.L., R.C., and G.K. aided in the experimental design and the collection of preliminary data. K.S. and Y.A. wrote the manuscript with input from all the authors. L.V.W. supervised the study.

**Competing interests:** L.V.W. has a financial interest in Microphotoacoustics, Inc., CalPACT, LLC, and Union Photoacoustic Technologies, Ltd., which, however, did not support this work. L.V.W. is a co-inventor of the patent titled "Transcranial photoacoustic/thermoacoustic tomography brain imaging informed by adjunct image data" (U.S. Patent No. 11,020,006), along with L.-M. Nie, X. Cai, K. Maslov, M. Anastasio, C. Huang, and R. W. Schoonover. This patent is relevant to the subject matter discussed in this manuscript. The rest of the authors declare that they have no competing interests.

**Data and materials availability:** The data that support the findings of this study are provided within the paper and its Supplementary materials. The reconstruction algorithm and data processing methods can be found in the paper. The reconstruction code is not publicly available because it is proprietary and may be used in licensed technologies.